# SURVEYING THE SOCIAL, SMART, AND CONVERGED TV LANDSCAPE: WHERE IS TELEVISION RESEARCH HEADED?

Marie-José MONTPETIT[1]
Research Laboratory of Electronics
Massachusetts Institute of Technology, Cambridge, MA, USA
mariejo@mit.edu

Pablo CESAR
CWI: Centrum Wiskunde & Informatica
Amsterdam, The Netherlands
p.s.cesar@cwi.nl

Maja MATIJASEVIC
Faculty of Electrical Engineering and Computing
University of Zagreb, Zagreb, Croatia
maja.matijasevic@fer.hr

Zhu LIU
AT&T Laboratories
Middletown, NJ, USA
zliu@research.att.com

Jon CROWCROFT
Computer Laboratory
Cambridge University, Cambridge UK
Jon.Crowcroft@cl.cam.ac.uk

Oscar MARTINEZ-BONASTRE
Department of Statistics, Mathematics and Informatics
Miguel Hernandez University, Elche, Spain
ombonastre@ieee.org

---

[1] Corresponding author

## Abstract

*"The TV is dead"* motto of just a few years ago has been replaced by the prospect of Internet Protocol (IP) television experiences over converged networks to become one of the great technology opportunities in the next few years. As an introduction to the Special Issue on Smart, Social and Converged Television, this extended editorial intends to review the current IP television landscape in its many realizations: operator-based, over-the-top, and user generated. We will address new services like social TV and recommendation engines, dissemination including new paradigms built on peer to peer and content centric networks, as well as the all important quality of experience that challenges services and networks alike. But we intend to go further than just review the existing work by proposing areas for the future of television research. These include strategies to provide services that are more efficient in network and energy usage while being socially engaging, novel services that will provide consumers with a broader choice of content and devices, and metrics that will enable operators and users alike to define the level of service they require or that they are ready to provide. These topics are addressed in this survey paper that attempts to create a unifying framework to link them all together. Not only is television not dead – it is well alive, thriving and fostering innovation and this paper will hopefully prove it.

# 1  Introduction

Television is now being redefined from a unidirectional flow of content from an operator to a device to a much richer combination of real time, on-demand, web and user generated content complementing traditional programming.

With more and more platforms (tablets, web enabled TV sets, smartphones, traditional set-top boxes, etc.) available for content distribution and consumption, TV is moving from traditional broadcast and multicast to a more personal video device ecosystem. TV programming is now moving from the traditional kind: it is geo-localized, time and place shifted and it combines video streaming to Web 2.0 messaging and widgets. And it is more and more wireless and mobile, with phenomenal growth predicted in the next few years [1].

This survey paper reviews the research challenges of the next generation television. While critics have declared that this whole field is now commoditized we intend to show video-rich data distribution is still an active area. Novel architectures for networks and middleware, improvements to the reliability of the connected information nodes, operations under reduced wireless availability, new distribution models to maximize the performance in wireless environments and the use of bottleneck resources, are all needed to provide the video-centric services of the future. We present a view of next generation television that leverages the benefits of recent networking and content distribution paradigms to build a truly *video centric network* approach to allow television to move *from network-TV to networked-TV*. By this we mean a TV network that uses the structure of information and the peering of wireless and wireline elements to provide video distribution with increased throughput, lower delays, better usage, and higher user satisfaction. We believe that the challenge of TV in the next decade necessitates a comprehensive end-to-end and top to bottom strategy that moves away from the current design silos. This is counter to still prevalent thinking. But the opportunities for smart, social and converged TV that combine video content and ancillary services like social commentary are need to move away from the *balkanization* of devices and networks.

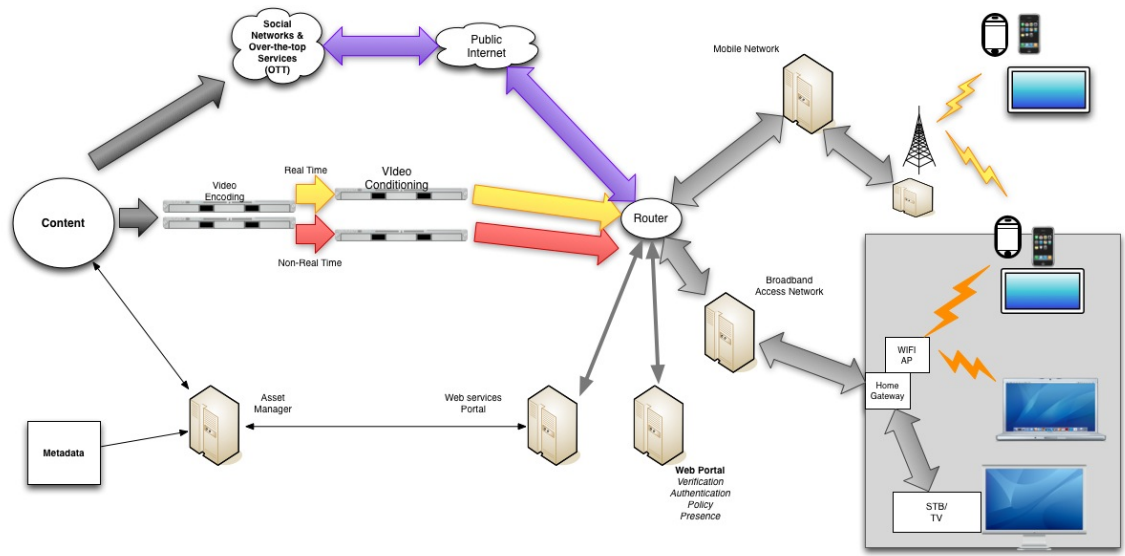

**Figure 1 Architecture Overview**

The paper will first define what we mean by converged and smart TV. Social TV which is one avatar of this convergence will also be presented with a view of the *beyond Twitter to the TV* of the first generation deployment and address the use of the multiscreen. New dissemination mechanisms are needed to ensure the distribution of the video content to applications and end users as well as the definition of basic networking architecture: these are two other sections of the paper. Of course, performance is the key to keep end users' loyalty. We will present novel approaches to ensure performance and the mechanisms to measure the quality of the experience as well as avenues for content and privacy protection (that are also addressed in a paper of the special issue). Finally, we will conclude the editorial with what we perceive are the challenges facing TV research as well as presenting the papers that form the bulk of the Special Issue.

# 2 Converged and Smart Television

Figure 1 presents a simple overview of the converged video distribution chain. We will define throughout this paper *converged television* as a television service offered on a diverse ecosystem of devices; and *smart television* as a television service augmented by ancillary services such as widgets or web content. Converged TV has also been coined *CE3.0* as it combines Internet with traditional consumer electronics [2]. We mentioned in the introduction that in the recent years video consumption has changed radically. We have seen the emergence of a more and more interlinked network of access networks and end-devices replacing the linear broadcast delivered to a single TV set of the recent past. The use of *social networks for video distribution and recommendation* [3] also figures prominently in this evolution. New services, in particular *social television* [4], require dramatic changes to the way video is delivered as they combine voice, video data and graphics [5][6]. The work is also at the forefront of standardization efforts in ETSI TISPAN, W3C, OpenIPTV and ATIS IIF, etc. and has influenced the directions of television research as well as highlighted some of the needs for better user experience and interaction, as well as improved transmission and network performance to provide the quality video experience users now demand over any network in and out of the home.

Converged television also aggregates and distributes multiple sources of television content onto different devices. While providing a simple user interface that masks this aggregation, converged television requires a reliable and comprehensive system and network architecture for content management and device interoperability. The converged television architecture allows applications to provide the services of a Common Distribution Network provider or a Repurposed Content Aggregator. Both models enable the distribution of content over any type of network, whether cable, wireless, or mobile phone, as well as to any connected device, especially over IP networks.

## 3   Social Television

*Social Television (Social TV)* has gone from the laboratory [7] to the boardroom with an incredibly swift pace over the last 2 years, a testament to the power of both television as a reflection of social trends but also to the popularity of social networks. Social TV was also rewarded with a MIT Technology Review TR10 in 2010, as one of the *influential 10 technologies* that will change the way video is consumed [8], and was the focus of an IEEE Networks feature [9].

Social television is reshaping the way people find and consume television content, providing research challenges for the multimedia research community. From the content modeling perspective, novel models are needed that take into account social interaction and that help in real-time manipulation (and combination) of television streams. Multimedia retrieval can benefit from current applications that are capable of aggregating and analyzing social network activity. At the system level, there is a need for robust platforms that assure QoE, including inter-destination synchronization of content streams (remotely joint watching). Finally, from the human-centered perspective, novel metrics for evaluating user satisfaction and engagement around TV content are needed. Such metrics should take into consideration sociability aspects.

Most Social Television implementations combine social interaction and personalization features, creating a user and community-centric viewing experience [10]. Social TV user interface design is aligned with work on contextual interfaces made popular by advanced smartphones and gaming systems, as it adapts to user behavior and preferred devices.

Given current media coverage [11], people might think of social TV as the result of aggregating social networking streams around television content. The reality is that the objective of original experiments in this field was to connect separated living rooms, creating a virtual living room where people could watch television together when apart [7]. Unfortunately, these experiments missed a golden opportunity, since they did not foresee the impact social networking has actually had. First classifications followed as well the initial model, where connecting living rooms was the key [12]. Nevertheless, more recent frameworks [13] take into consideration current social practices around television content (e.g., commenting, liking, status updates, and checking in).

But Social TV is, after all, just another step towards human-centered television [14][15][16] where the viewer becomes an active node in the television content flow, as producer, distributor and contributor. In this scenario, the user attains increased interactive capabilities within content streams for himself and for his social network. Independently of the reach and the size of such network (family, friends, the world), social TV enables users to interact around and within television content.

From a human-centered perspective, television content becomes the context of social interaction between people, allowing viewers to recommend (1), to annotate (2), to gather around (3), and to influence content (4). Each of these categories has a non-computer mediated counterpart as people recommend TV programs and movies to others, discuss last night's debate during lunch, go to brother's home to watch the Super Bowl, and vote off contestants in reality shows.

*Content recommendation* is probably the most popular application linked to social television (e.g., yap.TV[2] and Zeebox[3]), due to its straightforward business model. Advertisers and producers alike have recently discovered the value of social networking for better gathering (and sometimes influencing) people's interest. Freely available profiles are valuable assets for content producers, since they can use social interactions for providing more accurate content recommendations [17]. More interesting from a human-centered perspective are direct content recommendations, where users can share enriched video fragments.

According to Yahoo! and the Nielsen Company[4], 86% of mobile Internet users and 92% of the 13-24 youths are using their mobile devices simultaneously with TV, where updating/reading social networking sites is the second most popular activity. The second category of Social TV applications, *content annotation*, refers exactly to that, enabling viewers to comment on television streams. Some examples of this category include mobile applications such as Into_Now and Miso, second screen applications such as Zeebox and solutions integrated in the set-top box such as GoogleTV. However, some of these applications tend to create an information overload problem because microposts are not properly filtered and as a result, viewers are exposed to all the comments other people are making about a television program, which in most of the cases are irrelevant for them. Two interesting developments within this category are to filter microposts and to use social interactions for better describing television content. The Hulu on Facebook app[5] is an example of the former. By restricting the reach of the comments added to videos, viewers will not be overwhelmed by the amount of comments (only if people did not have around 130 friends on average[6]). An example of the latter is research intended to annotate television streams based on the activity of millions of users [18], similar to current trends in crowdsourcing for improving multimedia content retrieval.

The next category, viewers *gathering around content*, provides support so viewers in different locations can gather around television content. Allowing people to synchronously communicate with others while watching TV, these applications want to remediate social dislocation and to recreate the living room or water cooler experience in cyberspace. More recently, a number of novel applications are available. Some are extensions of instant messaging solutions, capable of embedding videos while chatting [19]. Some are virtual viewing rooms, where remote users can communicate with others, while watching television content together (e.g., YouTube Social, ClipSync, Starling), while others, such as Google TV, incorporate cameras and microphones following the model of virtual living rooms. In all these cases there are a number of challenges still to be solved. First, synchronization of the television stream across locations is still rudimentary, causing communication problems. Second, intelligent mechanisms for managing the conversation streams should be developed for enabling a pleasant experience (e.g., spatial audio). Finally, developers should be aware that not all genres allow for conversation [16], but some of them highly benefit from it, for example, quiz games where friends in different locations can compete, alongside the contestants.

---

[2] http://www.yap.tv/
[3] http://zeebox.com/
[4] http://advertising.yahoo.com/industry-knowledge/mobile-shopping-insight.html
[5] https://apps.facebook.com/huluapp/
[6] https://www.facebook.com/press/info.php?statistics

The last category, *influence the content*, is probably the most interesting and promising one. Unfortunately, it is the least developed due to the production costs. This category refers to provide viewers the opportunity to affect television content, while watching it. In this case, television viewers can (in an aggregated manner) socially influence what they are watching. Even though these concepts have created a new popular genre by allowing people to vote off contestants (e.g., Lost, American Idol), existing professional television productions are few. One exception was a television drama created in Finland, Accidental Loves, which viewers could influence in real-time by sending SMSs [20]. In this direction there is still much research to be done, since storytelling is a complicated task that cannot be taken lightly. Nevertheless, there is much to gain by involving content production companies on the goal of making television social, beyond checking how programs are doing on Twitter.

# 4 Advances in Content Dissemination

The provision of converged, smart and social television of the future creates other challenges for the networking community. We need to investigate *device composition* via novel peer-to-peer approaches and using the layered structure of new codecs, *network collaboration* by judiciously managing the end-to-end resources as well as novel approaches to *content discovery* and *user interfaces*.

In particular there is a need to get peer-to-peer (P2P) and super-distribution out of the realm of illegal distribution and into the mainstream, especially to leverage edge resources and the increase in CPU and memory capabilities of consumer electronics. There are opportunities to offer other services over the peer network by virtualizing the extra space or extra computing resources such as was proposed in the European Union Project Nanodata Centers[7]. Thus collaborating peers can be chosen because of proximity or availability of specific features like CPU or storage space. But more and more there is an interest in adding social network aspects to the choice of peers going beyond current mechanisms based on traditional device and service discovery using *social content discovery*. In addition there are more and more incentives to investigate the new field of *collaborating access networks or network combining* for *video aware* networking. This leverages the work that has been done in software radios and heterogeneous connectivity in Future Internets.

## 4.1 TV Content Analysis and Search

With the widespread adoption of IPTV and smart TV services, the amount of multimedia content that is available on consumers' TV screens becomes virtually unlimited. While it is desirable to have such enormous viewing choices, the need for a powerful and easy to use content search mechanism is essential. Effective content search also enables content owners and distributors to monitor and manage the content flow in a large scale TV service. Research in this area has been very active in the last two decades. TREC video retrieval evaluation (TRECVID) [21] is sponsored by the National Institute of Standard and Technology (NIST) to stimulate the video content analysis, indexing, and searching research. Considerable novel video processing systems and algorithms have been reported by TRECVID participants over the years, many of which can be directly applied to TV content. This section

---

[7] http://www.nanodatacenters.eu/

focuses on TV content analysis and search, as well as relevant standards.

### 4.1.1 TV Content Analysis

Content analysis and processing technique plays an important role in the success of TV content management system. After almost two decades of research in multimedia content processing, many innovative technologies have evolved nowadays to be commoditized.

From the perspective of involved modalities, TV content processing can be categorized into single-modality and multi-modality approaches. Single-modality methods rely on one type of information, including electrical programming guide (EPG), linguistic information, acoustic information, or visual information, while multimodal approaches utilize a combination of these cues. EPG provides metadata, including the TV show title, brief description, main casts, genre, etc. Linguistic information can be extracted from the accompanying closed caption / teletext stream, or a speech recognition engine. Acoustic based methods discover various audio events, for example, detecting the speaker change and classifying music and speech segments. Visual based approaches identify the content using visual features, which typically include color, edge, texture, etc. While single-modality methods analyze the content efficiently, multimodal approaches usually deliver higher performance and robustness due to the complementary information among different cues.

TV content processing methods can also be grouped by the addressed semantic levels. Content processing frameworks usually adopt a bottom up approach, where low level audio/visual/textual features are extracted at the bottom, and high level semantically meaningful content descriptors are determined at the top. Samples of the low-lever features include color, texture, edge, scale-invariant feature transform (SIFT) features in the visual domain and volume, zero crossing rate, and pitch features in the acoustic domain. Typical high level content descriptors are video summarization, video topics and concepts, etc. In the middle of this content processing pyramid, there exists the so called semantic gap, which represents the challenge in understanding the content based on low level multimedia features. Significant amount of research has been devoted to bridging this gap over the last decade.

In terms of the scope and applicability, TV content analysis can be either for general purpose or for specific domains. Usually, the processing at lower feature level can be applied in wide range of applications, while the high level processing that relies on more prior knowledge is tuned to individual applications. Certain TV programs, including news, sports, talk shows, weather forecast, etc., have well defined semantic structures. For example, news usually starts with the highlights introduced by the anchorpersons, and the detailed coverage from the reporters follows immediately. This rich prior knowledge makes the content analysis more tractable and effective. Other interesting research areas include commercial detection in TV and TV content retargeting. Detecting commercials is useful for the advertisers to monitor the commercial airtime and provide addressable advertisements to the end users, and it also allows viewers to locate interested commercials and skip the unwanted content. Video retargeting is to transform an existing video to fit other rendering devices with different display resolutions. This becomes particularly important due to the prevalence of consuming TV content on mobile devices and hand held tablets.

## 4.2 TV Content Search

The traditional way to find interesting TV content is by tuning to the Electronic Program Guide channel, and linearly scanning all listed programs on the screen. A more modern approach is the Interactive Program Guide (IPG), which allows the customers to navigate and browse the program information with a remote controller. IPG usually also allows the user to search TV program by metadata, including title, genre, rating, etc. Both interactive and the non-interactive EPG searches work fine when the number of TV channels is limited, but they quickly become ineffective when the volume of content experiences an exponential growth.

Most of the state of the art video search engines rely on content-based media processing techniques to provide a semantically meaningful representation of the indexed content. For example, speech recognition provides a linguistic description for the spoken content of videos and image analysis supplies richer content tagging. Content-based processing also enables value added services, including content recommendation and personalization services. In addition to the manually generated metadata of the viewed content, the content itself is employed to build the viewer's preference automatically and to make recommendations for future watching/recording intelligently.

Searching interfaces for TV programs evolve from text only to multimodal mechanism, where speech, gestures, and recorded audio/video clips are accepted. The relatively small form factor of the mobile devices and the wide availability of the multimedia sensors make them the perfect TV companion devices for content search and other non-viewing activities. Given the sweeping availability of smart mobile devices, TV content consumption on mobile devices has become the new trend.

Peer-to-Peer (P2P) networks have been considered to distribute large volume of TV content due to their intrinsic scalability and efficiency. Traditional centralized content search is extended to a distributed scheme in the P2P platform. Due to the higher than ever popularity of social network services, including Facebook, Twitter, and YouTube, TV content rating, recommendation, and mining based on social network data become more influential. Bluefin Labs[8] continuously monitors over 200 television channels in US, and sifts through more than 50 billion tweets and Facebook posts simultaneously. Analysis of such tremendous amount of information discovers the insights about TV shows and commercials, which are undoubtedly valuable for TV content search.

## 4.3 Standards for the TV Metadata

The MPEG-7 standard, formally named Multimedia Content Description Interface, provides a comprehensive set of audiovisual description tools for describing multimedia content. Although the scope of MPEG-7 is much broader than TV applications, it can be seamlessly applied in TV content index and search. With the importance and wide penetration of TV services, explicitly TV related standards have been developed too. Following are a few of them.

In ATSC digital television system, the Program and System Information Protocol (PSIP) [22] is used to carry metadata information. PSIP defines a set of tables for representing data in the MPEG-2 transport stream. The content ratings are carried in a Rating Region Time Table (RTT). The Event Information Table (EIT) contains content description, including start time, duration, title, and optional description, content advisory data, and metadata about the closed caption and audio (not the data itself). Descriptions may be sent using extended text messages (ETM) in extended text tables (ETT) and up to 16 days of program data may be advertised in advance.

Digital Video Broadcasting project (DVB) is an industry-led consortium that designs open technical standards for the delivery of digital television and data service. It has been widely adopted globally, especially in Europe, African, Australia, and Asia. Within this suite, the standard that is responsible for the metadata information is called DVB-SI (Service information) [23], which provides readable information about the TV content. Like the ATSC specifications, DVB uses MPEG-2 transport streams (TS) but the protocol for program metadata differs and is encoded in Program Specific Information (PSI) tables to include

---

[8] http://bluefinlabs.com/

service information. This is used for delivery of EPG information in DVB systems.

TV-Anywhere[9] is a set of specifications for the controlled delivery of multimedia content to a user's personal device (Personal Video Recorder). It is part of the European Telecommunication Standards Institute (ETSI). It seeks to exploit the evolution in convenient, high capacity storage of digital information to provide consumers with a highly personalized TV experience. Users will have access to content from a wide variety of sources, tailored to their needs and personal preferences.

The Alliance for Telecommunications Industry Solution (ATIS)[10] is a standard organization that develops standards and solutions for the telecommunications companies. The ATIS Interoperability IPTV Forum (IIF) defines the overall industry reference architecture and critical standards to support IPTV's effective deployment. Within ATIS IIF, the metadata and transaction delivery committee is responsible for defining metadata elements, the representation of metadata elements and the content of application level transaction. Recently, ATIS IIF released two metadata specifications: the IPTV Electronic Program Guide Metadata Specification and the IPTV Emergency Alert System Metadata Specification, as well as four metadata standards that provide the data structures to support content on demand.

## 5 Towards a Video Centric Network

New models of TV consumption must be met with specific end-to-end distribution architectures by taking into account for example the broadcast nature of the wireless medium and the *social nature* of the new video experiences. This is both an *evolutionary* solution, in terms of reusing some of the existing infrastructure in a new way, and a radical *clean-slate* approach that considers that networks *do not end at gateways*, devices can *operate beyond their shells*, and that video services are *not there to impose but to propose.* Many analysts have reported the emergence of the dominance of video traffic on today's Internet. Slowly replacing the old structured, and unstructured swarms of P2P video distribution systems, now centralized repositories like Youtube and the BBC use Content Distribution Networks, such as Akamai, to scale out delivery of the many petabytes of data to millions of viewers.

As discussed elsewhere in this editorial, viewing is now more often time shifted via DVRs, even in homes on the end of traditional broadcast channels. Live streaming is one fraction of traffic - of course it is an important category, since some of the most popular content is live coverage of sports events, and this is as social an experience as you can get (viewing on large screens in public places). However, the bulk of content is viewed at a different time from original publication.

In the early days of video on the Internet, this was not understood, and the community pushed for a network architecture that used multicast for global realtime multimedia delivery. This has other uses, most notably for many-to-many applications. However, the time-skewed viewing we have now moved to does not need simultaneous bulk delivery.

This means that network support should evolve towards supporting rendezvous between publisher and subscriber. This can be achieved in overlay networks, such as the aforesaid Akamai CDN or P2P systems like that built by Zattoo Networks. On the other hand, as we scale up the quantity and quality of the video, it becomes necessary to push such support down into the network layer itself. Hence, Future Internet Architecture considerations call for ideas like Content Centric Networking and software defined networks. In these systems, the support for matching interest to tags on content (or hashes of content name combined possibly with other attributes), is provided inside network layer routers, alongside (or replacing) IP.

---

[9] http://www.etsi.org/WebSite/Technologies/TVAnytime
[10] http://www.atis.org/

## 5.1 Using Principles from Content-Centric Networking

Content Centric Networking lends itself to various kinds of multi-tree routing, where content is retrieved from various places within the network, possibly including other end systems (i.e., it can subsume P2P networking). In today's access links, whether ADSL or 3G, we typically see limitations of uplink capacity which preclude sourcing a full video stream at full rate from an edge device. On the other hand, several devices in the home or the hand can together deliver a required rate for a live viewer. Alternatively, one can combine sourcing video from edge devices with super-peers (so-called peer assisted). The peer assist comes from better provisioned nodes just inside the network (basically, these evolve out of the old CDN servers). They can also help with management tasks like key distribution/ digital rights management (DRM) and payment/access control as needed; and practical workarounds for problems like NAT traversal. The software architecture for such networking leads to a more equal treatment of storage and networking in edge devices. It also provides more symmetry in the roles of all devices (the old end system versus intermediate system separation is elided). This allows clean separation of considerations of content delivery from rights management, freeing up the technology to evolve along two dimensions. Some challenges for this approach certainly remain.

Scaling the rendezvous mechanism to billions of end systems and channels is not trivial. There are different ways to map content, based on names or common channels, which interact with other parts of the architecture, especially subscription payment/rights and efficient key distribution.

The current Internet has many network layer players who have a set of interdomain agreements and settlements that have evolved over the last 20 years. It is not at all obvious how to map the future content distribution topology onto the network layer topology in a way that is consistent with sensible incentives and viably future business models for future TV content creators/publishers/subscribers, and network operators.

## 5.2 Digital Rights Management (DRM)

DRM is hard. DRM without lock-in is very hard. We need to avoid penalising users with poor resources near the outskirts of the net, versus high centrality devices. Incentives for sharing uplink capacity and resources, especially on mobile and battery powered devices. Providing unicast, end to end services of the traditional type (a phone call) and many-to-many realtime collaborative experiences now become a stretch, but are by no means impossible. However, the details need figuring out, especially aspects of synchronisation and privacy and integrity of content. The W3C and others are looking into innovative ways of providing protection for commercial content with minimal overhead as necessitated by wireless distribution of TV content.

Integrating IP based delivery with Digital Broadcast channels is essential (for highly popular content, broadcast will scale somewhat better for upwards of a thousand 24*7 channels, in capacity and energy terms). Policy tussles will occur right up to the set top box/home hub.

# 6 Quality of Experience

More and more, media consumers turn to IP-based TV content and are moving towards ubiquity. Without a doubt, wired high speed networks offer TV content with first-class guarantee of Quality of Service (QoS) metrics which traditionally has resulted in high user satisfaction. Mobile networks however, which are becoming prominent delivery channels for IP based content still require large infrastructural investments.

## 6.1 Performance Aspects

What does it mean to meet the challenges of heterogeneous devices and network and address the main quality impairments that are endemic to TV in heterogeneous networks? Video traffic meets variable bandwidth and SNR, intermittent connectivity, inconsistent dissemination delays, stateful implementations and finally, varied device ecosystem and quality requirements. Hence to provide the best experience necessitates to revisit networking decisions taken over a decade ago.

Recent advances in Network Coding aim to improve the performance of middleware elements in heterogeneous networks with multiple rendering devices. Random network coding has proven effective in optimizing network resource consumption in wireless networks and our work indicates that it can be applied across the whole OSI stack, from the waveform to the IP layer[24]. It has been applied to the problems of storage [25] and video downloading [26] that makes it ideal for television systems.

As was presented in previous sections, we have entered the age of the *TV viewing ecosystem*, where networks of smartphones, tablets, home gateways and set-top boxes (STBs) and digital video recorders (DVRs) can create community-based video distribution network. This in turn should efficiently use the enhanced middleware resources and optimize the use of content distribution networks (CDNs). We have shown in our work that network coding to middleware elements that transmit and process TV content is essential to these advances as if a node is down or moved out of reach due to mobility or an error, the files can still be reliably recovered [25]. This leads to policy development and the refinement of in home network management systems as well as improving the performance of peer-to-peer video distribution as was seen in previous sections.

## 6.2 User Behavior Effect: User Perceived QoE vs Network QoS

As mentioned earlier, subjective quality testing is very difficult and a large number of users are required for statistically relevant results. Expectations and prior experience, as well as individual motivation and personal sensibility to certain artifacts resulting from quality degradation, may play a decisive role in subjective evaluation. In view of that, research on QoE is still in its infancy and shows potential on a number of directions like mobile TV, 3DTV and ubiquitous TV services among others. QoE for these promising scenarios is lacking concrete *operationalizations* for the visual experience of content on small, sub-TV resolution and screens displaying transcoded TV content at low bitrates.

Subsequently, QoE for TV services should be personalized. Would a QoE threshold be shared among TV users in a specific situation? Does the QoE need to be addressed personally? This is the actual default assumption in QoE studies related with TV services. At the opposite end of the spectrum, this would call for clustering of users according to their QoE in order to deliver their expectations anywhere and anytime.

Consequently, a major challenge is sharing of watching experience (social TV) under these new frameworks with a wide variety of services and contents. It includes the identification of other relevant factors and the quantification of their influence on QoE. Some of them contain various human factors such as users' emotions, beliefs, references, perceptions, behaviors and accomplishments that occur before, during and after use of TV services. These elements are escalating the speed of solid development of smart and Social TV.

But whether or not a service would achieve success on the market depends ultimately on the users, their satisfaction and willingness to pay for the service. The recent paradigm shift from *quality of service* (QoS) to *quality of experience* (QoE) reflects this by bringing the user into focus. While in research community discussions still abound on the definition of QoE, QoS, and related terminology [27], and new definitions of QoE are emerging [28][29], we take the definition of ITU-T as a convenient starting point for discussion.[11]

There are several aspects worth further looking into:
- the notion that the *application or service is the subject of the assessment* (rather than the *system*);
- the notion that the *quality assessment is subjective*, and thus reflects the end-user's personal needs, preferences, prior experiences, and expectations (rather than an easily reproducible and stable quality measure);
- the notion that the *quality evaluation* also depends on factors "outside" the system, collectively referred to as the *context* (which can completely change the service acceptability, everything else being the same – for example, the price of the service).

Hence, as has been nicely pointed out [30] the traditional view of "quality" (as a fundamentally technical issue) must be broadened in other aspects, ranging from economical issues to human and social behavior, to consider the whole "communications ecosystem".

---

[11] Definition of Quality of Experience (QoE) from the ITU-T P.10/G.100 Amd.2) (2008): The overall acceptability of an application or service as perceived subjectively by the end user.
QoE includes the complete end-to-end system effects (client, terminal, network, service infrastructure etc.). The overall acceptability may be influenced by user expectations and context.

For the purposes of this editorial, it is interesting to explore these aspects from the point of view of QoE for social, smart, and converged TV, and present some new directions related to QoE research.

## 6.3 Service Effect: The Service as a Subject of QoE Assessment

Technical knowledge on how services are nowadays built is immense – and yet, the interrelations between technical issues and the user-centric, subjective assessment are just beginning to be understood. It is clear though, that the overall acceptability depends on many factors, again with many complex interrelations.

Somewhere "under the hood" of IP-based and Web-based "smart, social and converged TV" still lies the "plain old TV", built primarily upon audiovisual content. Consequently, the QoE evaluation for smart, social and converged TV as a service will also in great part be related to subjective assessment of quality of audio and video components and quality of synchronization. A lot of research efforts still focus in finding appropriate subjective metrics and testing methodologies for audio and video. Because the subjective tests are costly and time consuming, and – perhaps most importantly – cannot be performed in real time, the challenge is to design objective tests which could quickly and reliably assess the outcomes of subjective testing for audio and video. Voice, and to somewhat less extent, audio, have historically received much more attention than video, due to massive application in telephony and VoIP. Research in that area is thus more "mature", and it created some fundamental concepts and methodologies, such as the Mean Opinion Score (MOS), the E-model, and perceptual evaluation of speech quality (PESQ) – to mention just a few. QoE evaluation for video and audiovisual content, which is more relevant for TV, is more recent, and new standards are just emerging [31][32]. What also makes new "smart" and "social" TV significantly different, and more exciting from "old" linear programming TV, are broader context and more sophisticated ways to control the presentation and interact with the multimedia content. While significant progress has been achieved in that direction, recent results show that even when having good or promising metrics and tests for audio alone and video alone, it is not straightforward to calculate or estimate the collective effect, especially with respect to different tasks and given some other constraints [33][34][35]. Adding widgets and web content into the (TV) picture (pun intended) will only add to the challenge. How applicable and how useful will the existing metrics and methods for evaluating be, for example, web browsing usability or file download QoE, when observed in combination with TV content? In addition to video, web content and widgets on the same TV screen, new modalities are emerging, such as 3D TV and free-viewpoint TV (FTV), and along with them the respective QoE metrics. Judging by the success and projected growth in 3D networked games, this may create an additional boost for social TV, too.

## 6.4 System Effect: User QoE vs Network QoS

Considering that, in a converged network setting, smart and social TV is expected to be available "anywhere, anytime, on any device," it is clear that the "system effects" in the context of the above definition depend heavily on the infrastructure over which the service is provisioned and delivered. In this context, the "infrastructure" means the whole end-to-end "chain": the client software and the user device (terminal) jointly present a human interface to the service (i.e., corresponding data and processing facilities (server, cloud)), while the network provides end-to-end connectivity (the Internet, for example) and network access (wired/wireless, fixed, mobile, nomadic).

The first link in the chain is a combination of a client and a device, which jointly provide content and interaction capabilities to the user. In the context of smart and social TV, many new technologies and devices, from smartphones and tablets to interactive tabletops, open a completely new world of viewing experiences and interactions to be studied. The second link, network level QoS, typically represented through characteristics such as bandwidth, delay, jitter, and loss, is a well established concept, but, as noted before, it presents only a part of the picture, and moreover – if it is good enough – an invisible one, from the perspective of the user. The key issue – still – is how to specify how good (for the network operator) is "good enough" (for the user), in order to satisfy both the economic and the technical constraints. From the purpose of relating the quality as perceived by the user and the network level quality, two complementary views, termed the application aspect and the communication aspect, which have been posed in context of networked virtual environments, a decade ago [36], still remain for distributed and interactive services such as smart and social TV. The application view relates to the question: How does an interaction at the application (user) level affect the communication characteristics? The communication aspect relates to the question: How are events at the communication level reflected at the application level? While rather generic mappings and models can be found in standards (e.g. ITU-T G.1010, 3GPP TS 26944-900), the problem how to relate QoE and QoS for a specific, given service in a given context has been, and remains to be, an open area of research. European COST action QUALINET[12] is devoted to all aspects of QoE. Some novel ideas on generic mathematical relationships between QoE and QoS have been presented in [28] and [29], QoE monitoring and QoE-driven adaptations in networks are also being considered in [34].

The social aspect of converged TV brings a shift from individual viewing towards group interactions built around the joint experience, and yet another challenge, namely how to assess QoS and QoE in multiparty and multi-endpoint communication (both client/server and peer-to-peer). The use of multicast, among other technologies, is considered as being of particular interest for QoE improvement and better bandwidth utilization for users in the same geographical area, which is also a typical use case for social TV.

---

[12] http://www.qualinet.eu/

# 7 Conclusion

In this paper we have summarized the research challenges of the next generation television. We conclude with two forward-looking statements.

First, we postulate that a key challenge in the next decade will be to follow a comprehensive strategy to end-to-end and top-to-bottom systems to move away from the current silos, and to encourage innovation across the viewing ecosystem. We refer to this as the device and service convergence challenge. Second, we emphasize the essential roles of smart, connected and social TV for combining video content, transmedia (and cross-media) and social commentary; this is refer as the content convergence challenge. Both of these challenges will be critical in the creation of the next generation TV experiences.

Hence, by reviewing the the main topics of television research today we are reminded that far from being dead television is more alive than ever. The changes to content consumption of all kinds in the past few years, has moved television from the *small screen* to the *any screen*. This in turn created a vibrant and multidisciplinary research community that addresses the move to television anywhere. By reporting on recent advances in television research this paper wanted to open more avenues for innovation and further investigations in the new television landscape.

# 8 Acknowledgements

Pablo Cesar's work has funding support from the European Community's Seventh Framework Programme (FP7/2007-2013) under grant agreement no. ICT-2007-214793. M.J. Montpetit's work was partly funded by NBC Universal. M. Matijasevic's work is supported by the national research project no. 036-0362027-1639 funded by the Ministry of Science, Education and Sports of Republic of Croatia. Oscar Bonastre's work has funding support from the Ministry of Science (Spain) under grant agreement no. MTM2008-06778-C02-01. Other acknowledgements include Henry Holtzman, Muriel Médard and David Geerts.

# Biographies

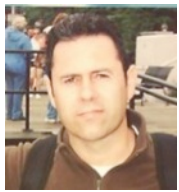
**Oscar MARTINEZ-BONASTRE** is an Assistant Professor and deputy director of International Relations at the Miguel Hernandez University of Elx (Spain). He received his Ph.D. in Telecommunication Engineering at the Polytechnic University of Valencia, Spain. His research interests include the development of congestion control, of reliable and layered multicast techniques for multimedia and of new solutions for IPTV and Social TV content distribution. Dr. Bonastre is a Senior Member of the IEEE.

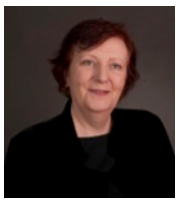
**Marie-José MONTPETIT** is a researcher in the Research Laboratory of Electronics at MIT and lecturer on Social TV at the MIT Media Laboratory. Her research is on novel approached to content dissemination especially converged video and social media. Her work has gained her international recognition, a Motorola Innovation Prize and a MIT Technology Review TR10. Dr. Montpetit received a Ph.D. in EECS from the Ecole Polytechnique in Montreal, Canada. Dr. Montpetit is a Senior Member of the IEEE and a member of ACM.

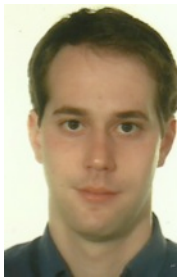
**Pablo CESAR** is a researcher at CWI (The National Research Institute for Mathematics and Computer Science in the Netherlands). He received his PhD from the Helsinki University of Technology in 2006. His research interest include multimedia systems and infrastructures, social media sharing, interactive media, multimedia content modeling, and user interaction He is involved in standardization activities and has been active in a number of European projects as well as a tutorial instructor in ACM and WWW conferences. He is co-editor of the book "Social Interactive Television: Immersive Shared Experiences and Perspectives".

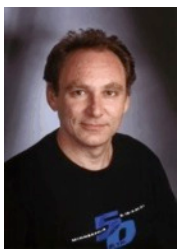
**John CROWCROFT** has been the Marconi Professor of Communications Systems in the Computer Laboratory since October 2001. His current active research areas are Opportunistic Communications, Social Networks, and techniques and algorithms to scale infrastructure-free mobile systems. He graduated in Physics from Trinity College, University of Cambridge in 1979, gained an MSc in Computing in 1981 and PhD in 1993, both from UCL. He is a Fellow of the ACM, a Fellow of the British Computer Society, a Fellow of the IET and the Royal Academy of Engineering and a Fellow of the IEEE.

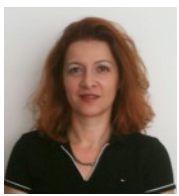
**Maja MATIJASEVIC** is a Professor in the Faculty of Electrical Engineering and Computing at the University of Zagreb, in Croatia. Her research interests include networked multimedia and quality of service in IP-based next generation networks. She received her Dipl.-Ing, M.S. and Ph.D. degrees in Electrical Engineering from the University of Zagreb and the M.Sc. in Computer Engineering from the University of Louisiana at Lafayette, LA, USA. She is a Senior Member of the IEEE and a member of ACM.

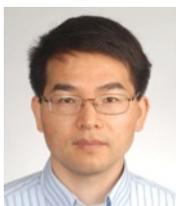
**Zhu LIU** received B.S. and M.S. degrees in Electronic Engineering from Tsinghua University, China and a Ph.D. in Electrical Engineering from the Polytechnic University, Brooklyn, NY. He is a Principal Member of Technical Staff in the Video and Multimedia Technologies and Services Research Department at AT&T and an adjunct professor of Electrical Engineering at Columbia. His research interests include multimedia content analysis and databases, video search, pattern recognition, machine learning and natural language understanding. Dr. Liu is on the editorial boards of the IEEE Transaction on Multimedia and the Peer-to-peer Networking and Applications Journal. He is a Senior member of the IEEE and a member of ACM.